\documentclass [twocolumn]{article}

\usepackage[textwidth=18.3cm, textheight=23cm]{geometry}
\usepackage{graphicx}
\usepackage{authblk}
\usepackage{secdot}
\usepackage{caption}
\usepackage{wrapfig}

\usepackage{amsmath}

\usepackage{titlesec}

\usepackage{newtxtext, newtxmath} 
\usepackage{parskip}

\titleformat*{\section}{\normalsize\bfseries}
\titleformat*{\subsection}{\normalsize\bfseries}

\begin{document}

\title{Voltage Ride-Through in Large Loads- A Dual $PQ$ Approach}
\author[1,2]{Amir Norouzi~(Member, IEEE),}
\author[1]{Michael Morel}
\date{}
\affil[1]{\footnotesize{Lancium, Newport Beach, CA}}
\affil[2]{\footnotesize{Corresponding Author: \texttt{amir[dot]norouzi[at]lancium[dot]com}}}

\maketitle
\textbf{
\textit{Abstract-}This paper provides a detailed investigation of voltage ride-through in large loads, such as Artificial Intelligence data centers. Voltage ride-through capability of large loads during transient disturbances in the power grid is important because of the potential impact on the stability and reliability of the Bulk Power System. A mathematical analysis is presented and it is shown how the traditional approach, based on reactive power compensation, may not be adequate for voltage ride-through in large loads. Ultimately, due to capacity limits of the load's power distribution infrastructure and grid's constraints, there is a limit to using reactive power as a corrective tool. A new dual active and reactive power (\emph{PQ}) approach is proposed in which non-grid resources with dynamic \emph{P} and \emph{Q} capabilities are shown to be needed to help with voltage ride-through. Additionally, the analysis illustrates that at extreme voltage dips in the power grid maintaining an acceptable level of load voltage can become practically or theoretically unattainable, which may lead to the load's disconnection from the grid. Analytical results are provided with practical numerical examples.}\\[0.2cm]

\section{Introduction}
\label{sec:introduction}
ARTIFICIAL Intelligence (AI) data centers have introduced a new category of electric loads that is unprecedented in size, density, and complexity. With a single computing rack currently sized at tens of kilowatts and increasing AI data centers are already surpassing 1GW of total load. Loads at this size are often energized by connecting to the power grid at the transmission level, or what is known as Bulk Power System (BPS). Any single load at gigawatt scale on the BPS can potentially impact the reliability and stability of the power grid, a phenomenon that has opened a new field of investigation.
System frequency and voltage are two critical indicators of stability or health of the power grid, and AI data centers, as large loads, can potentially affect both of these measures. An emerging and important topic stemming from the massive load size of the AI data centers is their capability of voltage ride-through during power grid transient disturbances, and how this can impact the reliability and stability of the BPS.

\subsection{\emph {Voltage Ride-Through in Large Loads}}
 The concept of ``ride-through'' has been developed for generating resources with the goal of supporting the BPS during voltage or frequency disturbances. In dealing with Distributed Energy Resources (DER) IEEE 1547-2018~\cite{1547} defines ride-through as the ``ability to withstand voltage or frequency disturbances inside defined limits and to continue operation as specified''. North American Electric Reliability Corporation's~(NERC) PRC-024-4~\cite{prc24} specifies frequency and voltage protection settings for synchronous generators, some wind resources, and synchronous condensers so as not to trip during some defined excursions for the purpose of supporting the BPS. Similarly, NERC's PRC-029-1~\cite{prc29} provides voltage and frequency ride-through requirements for inverter-based resources (IBR).

Large loads have a potential to similarly play a role in supporting the BPS during disturbances. In a recent report~\cite{nerc} NERC defines a large load as a load facility at a single site ``that can pose reliability risks to the BPS due to its demand, operational characteristics or other factors''. The basic idea is that when large loads disconnect from the power grid during disturbances, e.g., by switching to their emergency resources, the amount of load removed from the grid may become large enough that it can impact grid's stability. Hence, the need for large loads to be able to ride through grid disturbances within defined limits with the same goal that led to the development of ride-through requirements for generating resources, i.e., supporting the BPS for greater stability and reliability.

\begin{figure}[h]
\centerline{\includegraphics[width=3.5in]{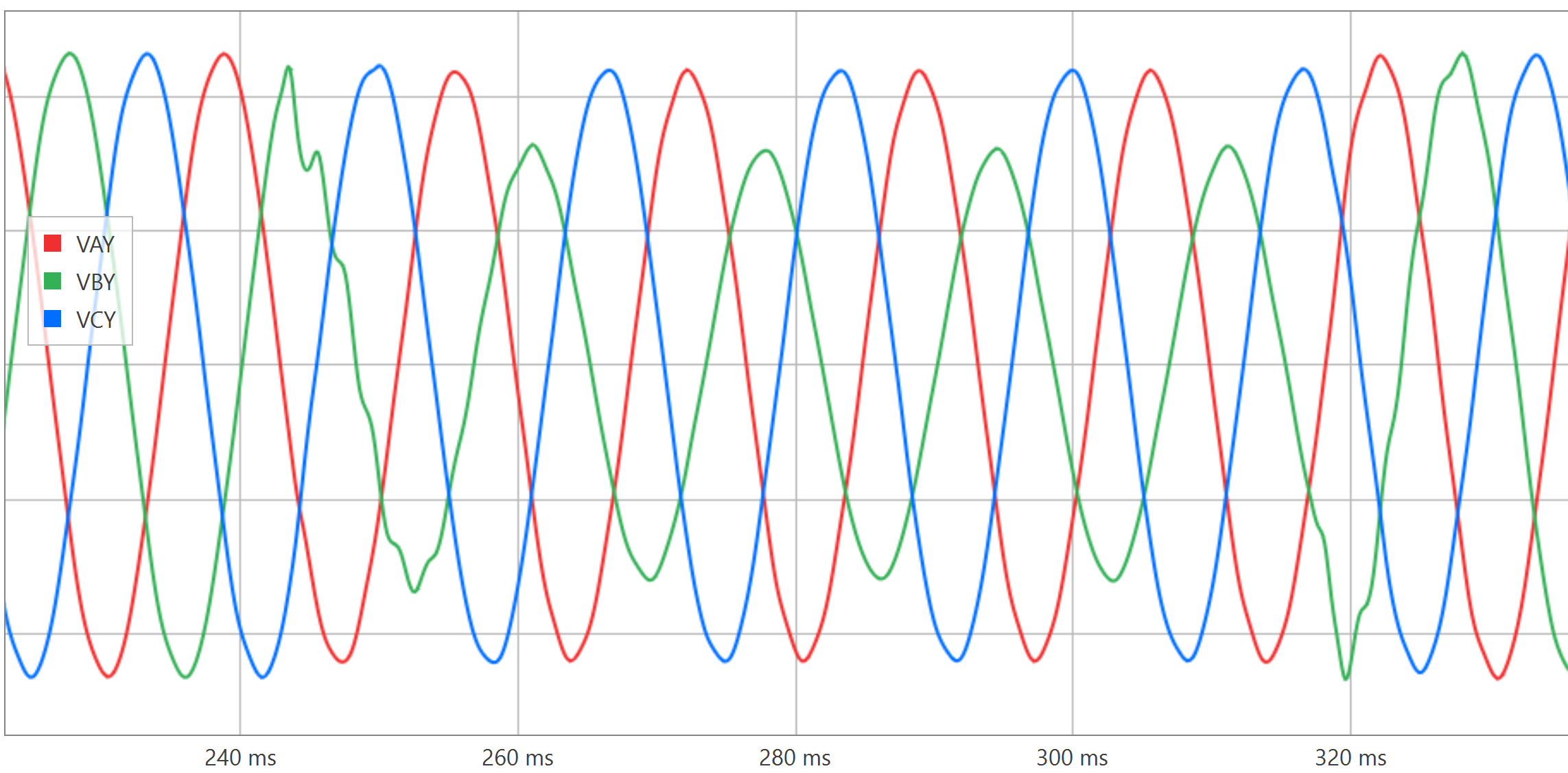}}
\caption{A Real Case of Grid Disturbance and Voltage Dip.
\label{fig:vdip}}
\end{figure}

A major source of transient disturbances in the power grid are faults, mostly as ground faults. Many of these faults are temporary in nature and generally last under 200ms, such as a flash-over across an insulator. Others are typically cleared by the protection systems within a similar time frame. Within the grid's stable operating conditions faults generally cause voltage disturbances, appearing as voltage dips where voltage magnitude temporarily drops. Depending on the severity of the fault and distance from the measurement location, voltage dips could be as large as $50\%$ of the nominal value or even larger. Figure~\ref{fig:vdip} shows a real-world voltage dip at a load, likely attributed to a fault on the grid.

The majority of load in AI data centers consists of computing hardware. For reliability purposes, computing loads are often powered through Uninterruptible Power Supply (UPS) units~\cite{chen}. UPS units have limited battery resources as emergency source of power and are designed to switch to these sources upon some defined disturbances on the power grid. From the perspective of the grid, switching of a computing load to non-grid resources is loss of that load, and depending on the size of the lost load as well as the grid stability limits this could cause other cascading events on the grid. This is why the capability of large loads to ride through voltage disturbances, most often due to faults on the grid, is emerging as a significant matter, particularly in AI data centers.

\subsection{\emph {Outline of the Research}}
There are major gaps in academic research as well as in the industry standards in understanding and addressing the problems related to the scale, dynamics, load behavior, operation principles, modeling, impacts on the grid’s reliability and stability, and disturbance ride-through of the fast-emerging AI data centers. It appears that currently the industry, including data centers and power systems, is leading the research due to practical needs. NERC’s Large Load Working Group has been recently established to identify the knowledge gaps in large loads and define relevant operation expectations and standards for grid’s reliability and stability, as can be seen in~\cite{nerc}. In~\cite{MS} some of the major AI technology leaders discuss operation principles of AI training from the perspective of electric power and the impact on the grid, proposing hardware and software mitigation methods.  Electric Reliability Council of Texas (ERCOT) with its Large Load Working Group is a prominent example of the power industry’s efforts to understand and ultimately provide voltage ride-through standards for AI data centers~\cite{ercot}. In a white paper~\cite{i3esa} published by the IEEE's newly-launched \emph{Data Centers: Standards Needs Analysis and Recommendations Activity} an extensive survey of the ongoing industry initiatives on data centers is provided. The report reviews various efforts underway in the power industry related to standards and grid readiness for data center deployment, including those by grid operators and utilities in North America and other parts of the world. In academic research, \cite{uab} was an early publication that provided descriptions of AI load characteristics and the impact of power electronic systems within the power hardware chain of large loads. In~\cite{chen}, a more recent publication, power infrastructure of AI data centers and the associated challenges are reviewed, including electricity markets, real-time dynamics, ride-through and stability, and power fluctuations. Addressing the significant size and energy resource needs of the emerging AI data centers, \cite{bri} provides a comprehensive analytical approach for siting of these large loads.

However, what is missing in the research is a detailed analysis of voltage ride-through, including rigorous mathematical investigation as well as analytical examination of solutions that may or may not be effective. The presented research in the industry and academia so far has been almost entirely descriptive accounts of the problem of voltage ride-through, or lack thereof, based on observations and known events, without much, if any, analytical studies. Besides, the investigations of voltage ride-through capabilities and expectations of large loads have been mostly based on ad hoc simulations, quite often not leading to an in-depth and coherent understanding of the problem. The lack of analytical studies of voltage ride-through in large loads has hindered the efforts to establish industry-wide standards and expectations for ride-through capabilities of large loads, including AI data centers, giving rise to confusion among both the data center industry and the electric utilities.

To address these research gaps this paper first provides an in-depth mathematical analysis of the theory of the interactions between active power, voltage, and reactive power from the perspective of loads in Section II, using the classical circle diagrams. In the classical theory of load power flow the role of reactive power ($Q$) is traditionally studied in the context of normal-state power factor correction, mainly for the purpose of improving voltage drop, within low single-digit percentages, across transmission and distribution systems. A novelty of the presented technical analysis is in its new perspective, which studies the role of $Q$ in maintaining an acceptable \emph{transient} load voltage during voltage dips of double-digit percentage on the grid, mostly due to \emph{faults}. In other words, the role of $Q$ in this new application is studied during \emph{transient} conditions as opposed to the conventional power factor correction during normal operating conditions. Next, theoretical and practical limitations of using $Q$ for voltage ride-through is rigorously determined, and then it is shown, in Section III, why and how active power may provide additional support for the problem of voltage ride-through. While reactive power is traditionally linked to, and used for, voltage regulation a new concept is developed which involves using \emph{active power}, in addition to reactive power, for voltage compensation and ride-through during grid faults or other transients. The analytical results are also further reviewed by numerical examples.

\section{The Interplay of Power, Voltage and Reactive Power in Loads}
\begin{figure}
\centerline{\includegraphics[width=2.5in]{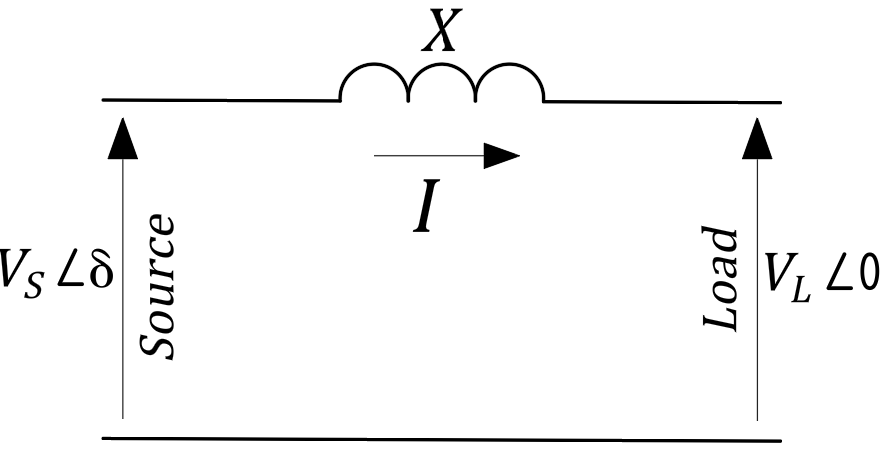}}
\caption{Representation of a Substation Serving a Large Load.
\label{fig:sld}}
\end{figure}

Since a fault-driven disturbance on the power grid often involves a dip in the voltage a good understanding of how voltage, active power ($P$), and reactive power ($Q$) are interrelated in loads is essential. Figure~\ref{fig:sld} shows a power system with the source voltage, $V_{S}$, the voltage at the load, $V_{L}$, and an equivalent system reactance, $X$. In large loads, such as AI data centers, $V_{S}$ may be considered the grid voltage at the load's substation's point of interconnection, most often $230 KV$ or higher, while $V_{L}$ may be considered the medium voltage at the main MV distribution facility, after the grid voltage is stepped down by the substation main power transformers. The equivalent impedance between these HV and MV points at the substation is almost entirely due to the main power transformers, each typically sized at tens of megawatts or higher. In these large power transformers the equivalent resistance is much smaller than the reactance and therefore is often ignored~\cite{fitz}. For example, in IEEE C37.010-2016 standard~\cite{c37} the average or typical $\frac{X}{R}$ for 100MW transformers is found to be about 33, which means that the transformer's resistance is only $3\%$ of its reactance, hence representing the substation impedance by a reactance, $X$.

The active and reactive power at the load side~(receiving end) in Figure~\ref{fig:sld} is governed by the following well-known equations~\cite{steve}:
\begin{align}
P&=\frac{|V_{S}||V_{L}|}{X}\sin{\delta}\label{eqP}\\
Q&=\frac{|V_{L}|}{X}\left(|V_{S}|\cos{\delta}-|V_{L}|\right)\label{eqQ}
\end{align}

These equations provide $P$ and $Q$ on the load side as a function of magnitude of source and load voltages, $|V_{S}|$ and $|V_{L}|$, as well as the phase angle between them, $\delta$. If voltages are the phase-neutral values $P$ and $Q$ must be multiplied by 3 for total three-phase values, but if the voltages are line-to-line the equations will directly provide three phase powers. We observe that direction of the power flow is independent of the absolute or relative magnitudes of the two voltages and is only a function of the phase angle difference, $\delta$, i.e., as long as $V_{S}$ is ahead of $V_{L}$(positive $\delta$) power flows from source to load, irrespective of magnitude of the voltages.

On the other hand, the sign and magnitude of the reactive power is a function of voltage magnitude and $\delta$. For a constant load voltage, $Q$ can vary in sign and magnitude with $|V_{S}|$ and $\delta$. Since in practice $\delta$ is small during normal operating conditions, and hence variations of $\cos{\delta}$ is not significant, $Q$ is mostly a function of magnitudes of the voltages.

It should be noted that the grid's strength, i.e., its equivalent impedance, impacts the interaction between the large load and the grid. However, that impact is already captured by $|V_{S}|$ in equations (\ref{eqP}) and (\ref{eqQ}), as it is the voltage at the point of interconnection between the load's substation and the grid.

\begin{figure}
\centerline{\includegraphics[width=3.5in]{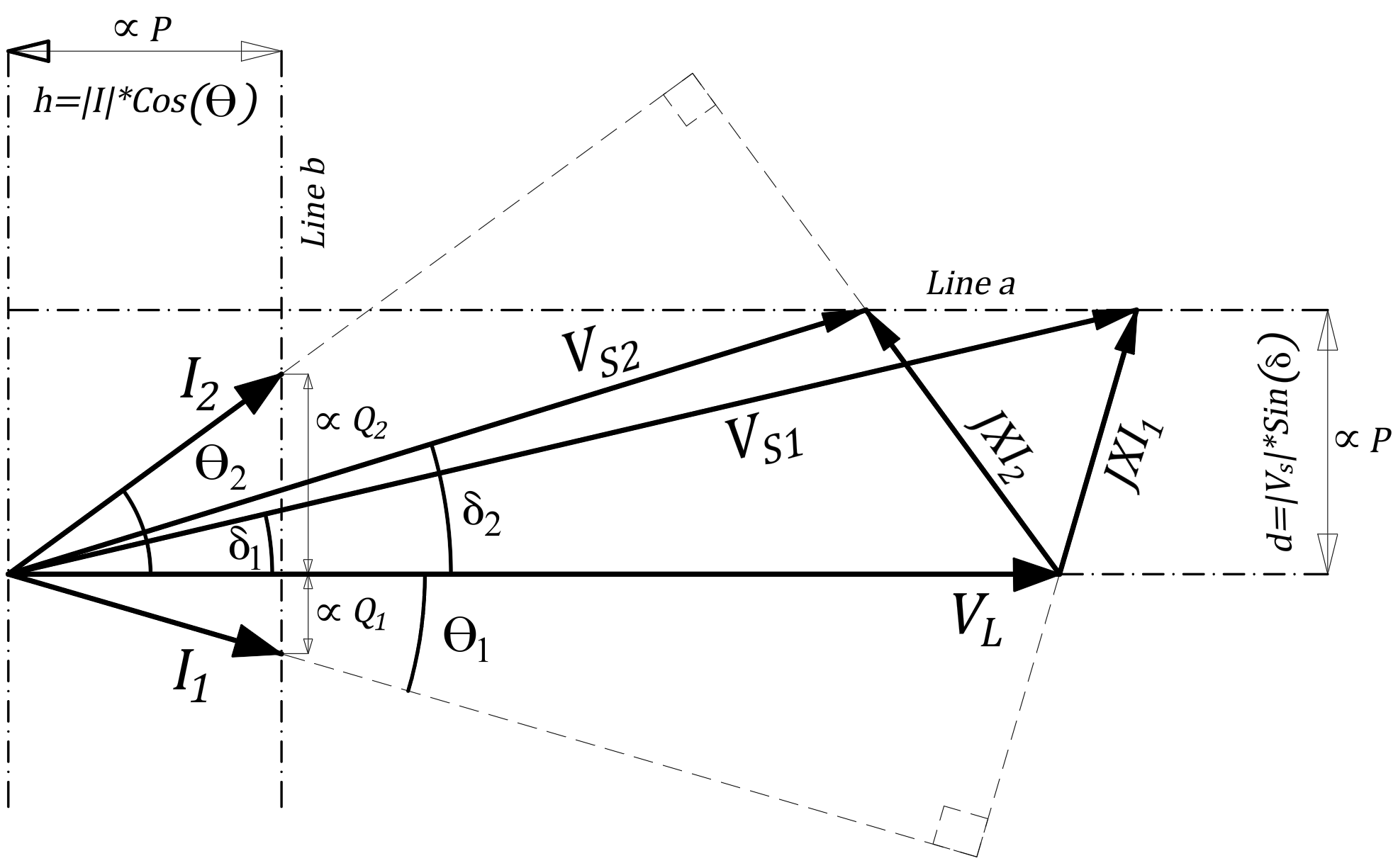}}
\caption{Vector Diagram for Figure~\ref{fig:sld} at Constant $V_{L}$ and $P$.
\label{fig:vector}}
\end{figure}

For the system in Figure~\ref{fig:sld} we have $V_{S}=V_{L}+jXI$. For easier visual representation and understanding vector diagram is used for this analysis. Figure~\ref{fig:vector} shows the vector diagram of this system for two cases: one with a slightly lagging load current and the other with a leading current, while load voltage and power are held \emph{constant}. Note that for a constant $V_{L}$ the vertical distance of $|V_{S}|\sin{\delta}$ is proportional to $P$, and that for any $V_{S}$ to transfer the same power $P$ this vertical distance must remain the same. Since $P$ is also proportional to $I\cos{\theta}$, where $\theta$ is the phase angle between $V_{L}$ and $I$, for any constant $V_{L}$ and $P$ the horizontal distance $I\cos{\theta}$ must also remain the same. Besides, note that for a constant $V_{L}$ the vertical distance $I\sin{\theta}$ is proportional to $Q$. A lagging current corresponds to $I\sin{\theta}$ being in the area below $V_{L}$ (positive or inductive $Q$) while a leading current corresponds to $I\sin{\theta}$ being in the area above $V_{L}$ (negative or capacitive $Q$).

In Figure~\ref{fig:vector} we notice that not only the magnitude of $Q$ but also its sign changes when current moves from $I_{1}$ to $I_{2}$ while the \emph{same power} is delivered to the load. This means that it is possible to deliver the same power $P$ at constant load voltage with different values of $|V_{S}|$, and this will require different values of reactive power, $Q$.

The vertical distance $d$ is equal to $|V_{S}|\sin{\delta}$ which is proportional to $P$ at constant load voltage, i.e., as long as $V_{S}$ slides on the horizontal Line $a$ the same power will be delivered to the load. When $V_{S}$ moves along Line $a$ from the far right to the far left $\delta$ changes from very small all the way to $90^{\circ}$, at which point $V_{S}$ has its theoretical minimum magnitude, equal to the distance $d$; furthermore, the horizontal distance $h$ is also proportional to $P$ at constant $|V_{L}|$ and while $V_{S}$ moves from right to left, current angle, $\theta$, moves from $-90^{\circ}$ to $+90^{\circ}$, sliding on the vertical Line $b$ for the constant power. The reactive power, $Q$, will move along the vertical Line $b$, changing magnitude and sign from reactive to capacitive accordingly. It is notable that while the current vector slides on the vertical Line $b$ its magnitude becomes larger and larger as the angle $\theta$ moves away from zero on the positive or negative sides.

Therefore, it is theoretically possible to deliver the \emph{same} power at \emph{constant} load voltage with a wide range of $|V_{S}|$, from very small to very large. This will require the magnitude of current and its angle, and hence \emph{reactive power}, to adjust accordingly. This is how $Q$ plays an important role in the interplay of $P$ and $V_{S}$. It is well-known that the lowest current to serve a load occurs when voltage and current are in phase, i.e., at unity power factor, and that is the goal of the traditional power factor correction. It involves making the load current in phase with voltage by providing appropriate amount, and type, of reactive power at the load bus.

\subsection{\emph {Anatomy of a Power Flow: The Circle Diagram}}
Equations (\ref{eqP}) and (\ref{eqQ}) can be used to examine the dynamics of power flow in more details. In equation (\ref{eqQ}), if we move the term $\left(-\frac{|V_{L}|^2}{X}\right)$ to the left side, then square both sides of equation (\ref{eqP}) and the manipulated equation (\ref{eqQ}), and finally add the two squared equations together we will arrive at the following equation, in which $\delta$ is eliminated:
\begin{equation}
P^2+\left(Q+\frac{|V_{L}|^2}{X}\right)^2=\frac{|V_{S}|^2|V_{L}|^2}{X^2} \label{eq:eqcirc}
\end{equation}
On a $PQ$ plane, equation (\ref{eq:eqcirc}) is a circle with its center at $(0, -\frac{|V_{L}|^2}{X})$ and with a radius of $\frac{|V_{S}||V_{L}|}{X}$. Note that the center is independent from $V_{S}$. For a known system $X$ and with a \emph{constant} load voltage, $V_{L}$, we can plot various circles, each corresponding to a certain $|V_{S}|$. For stability reasons, $\delta$ must not exceed $90^\circ$ and therefore the operation area is within the first quarter of the $PQ$ plane.

\begin{figure*} [t!]
\centerline{\includegraphics[width=4.2in]{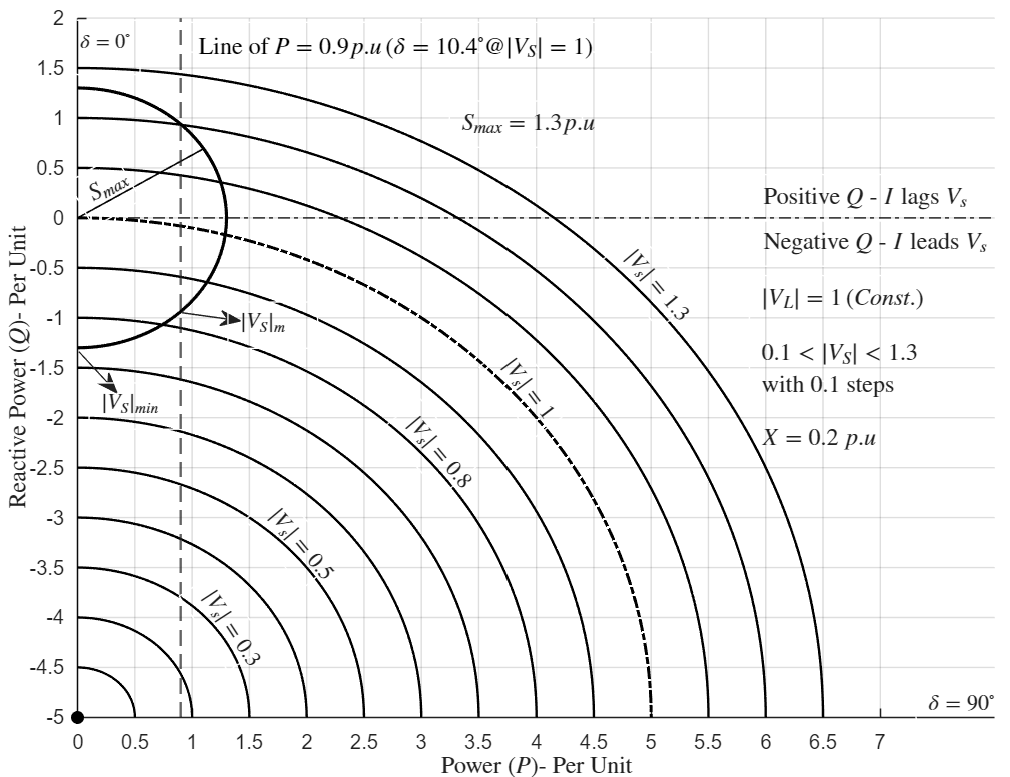}}
\caption{Load Side Power Circles at Constant Load Voltage.}
\label{fig:pcircle}
\end{figure*}

Figure~\ref{fig:pcircle} shows power circle diagrams on the load side or receiving end~(the circle with the radius of $S_{max}$ will be discussed later), shown only in the first quarter of the $PQ$ plane, produced in MATLAB, for a system with $X=0.20\hspace{0.75mm}p.u$, a constant load voltage of $V_{L}=1\angle0\hspace{0.75mm}p.u$, and $0.1\le|V_{S}|\le1.3\hspace{0.75mm}p.u$. Each plot corresponds to $0^\circ\le\delta\le90^{\circ}$, although in practice $\delta$ is very often below $30^\circ$. The values of $P$ and $Q$ are in per unit. The center of the circles is at $(0, -5)$. The largest circle is for $|V_{S}|=1.3\hspace{0.75mm}p.u$ and the smallest one is for  $|V_{S}|=0.1\hspace{0.75mm}p.u$. The theoretical peak of $P$ occurs at $\delta=90^\circ$, after which the system is considered unstable as the power drops with increasing $\delta$. Power circle diagrams are extremely helpful and were widely used in transmission line planning and performance evaluation~\cite{steve}\cite{saadat}; the widespread application of digital computers in power system analysis has significantly taken attentions away from this powerful tool~\cite{siman}.

We observe that for any $|V_{S}|\le1$ reactive power is negative, which, by convention, means current leads the voltage and that the grid absorbs $Q$ while the load supplies $Q$ at the same time~\cite{steve2}. In general, when $|V_{S}|\le1$ the load current must be capacitive or leading to support a unity load voltage. On the other hand, when $|V_{S}|>1$ reactive power could be positive (inductive), depending on the magnitude of $P$. At higher values of $P$ a negative (capacitive) $Q$ is needed to support a unity load voltage.

A vertical line corresponds to a constant power and shows what amount of $Q$ is needed at different values of $|V_{S}|$ to maintain a constant $|V_{L}|$ at the load, if possible at all. For example, for $P=0.9$ it can be seen that more $Q$ is needed at lower values of $|V_{S}|$, up to the point where $\delta=90^\circ$, which is the intersection of the $P$ line and the line of $\delta=90^\circ$, beyond which the system becomes unstable. This minimum $|V_{S}|$ can be calculated from equation (\ref{eqP}) at unity load voltage, known load power and system's $X$, with $\delta$ at $90^\circ$. For $P=0.9~p.u$ and $X=0.2~p.u$ the low limit of the source voltage is calculated as $0.18~p.u$, as can be verified in Figure~\ref{fig:pcircle}. Therefore, there is a theoretical limit to the amount of dip in $|V_{S}|$ where the load power can still be met at a constant load voltage by adjusting the reactive power, $Q$.

It should be noted that the analysis of the dynamic relationship between $P$, $Q$, and $V$, leading to the circle diagrams, while implying a consistent value for the reactance of the load's transformer, $X$, presumes \emph{stable} grid conditions, including during transients and faults. This means that faults and other transients are assumed to be cleared within the stability limits of the grid, where frequency of the power system remains stable and operationally acceptable. Some of the main conditions that cause core saturation, under which the reactance of a transformer, and hence its $\frac{X}{R}$, can substantially alter include high through-fault current, overvoltage, and large frequency deviations. Undervoltage condition alone does not contribute to core saturation of a transformer, unless it leads to an extremely high through-current. Of the mentioned three conditions, high overvoltages and frequency deviations are unlikely to occur under grid's stable conditions where voltage ride-through is studied or applicable. High through-fault currents that can saturate core of a transformer may arise either due to substation internal faults, or due to grid faults very close to the large load's substation. In the former case of substation internal faults voltage ride through is not applicable or possible, and in the latter case the voltage dip is likely going to be so large that voltage ride-through is again not plausible at all. The likely scenario that voltage ride-through in a large load can be achieved is when the fault or transient on the grid is electrically far enough that there is no significant fault contribution from the large load facility, and hence the risk of transformer core saturation is reasonably low, where ultimately its reactance remains mostly stable and consistent, all the while there is a notable voltage dip. It is also important to take into account that the on-site power resources of grid-connected large loads are either typically inverter-based resources, with limited ability for fault contribution, or in the case of generation based on conventional rotating machines, the installed capacity is often considerably smaller than the total load, which reduces the relative fault current capability of those machines. Overall, this is the justification for the premise that the transformer's reactance is unchanged, and hence circle diagrams remain applicable, under the conditions that voltage ride-through is reasonably expected to be achieved.
\subsection{\emph {Reactive Power and Voltage: Conceptual and Practical Constraints}}
The goal of the conventional reactive power compensation is power factor correction, i.e., making the load power factor as close to unity as possible. At a unity power factor power can be delivered at the smallest possible current, hence the smallest voltage drop across the transmission line which helps keep the load voltage within an acceptable level.

With load voltage ride-through we are dealing with a different problem than the conventional reactive power compensation. As we saw earlier it is theoretically possible to provide the same power, at a constant load voltage, by different levels of source voltage, $|V_{S}|$. For the case of a dip in $|V_{S}|$, i.e., a reduced source voltage, a leading current with a larger magnitude may be able to meet the goal.

\begin{figure} [t]  
\centerline{\includegraphics[width=3.5in]{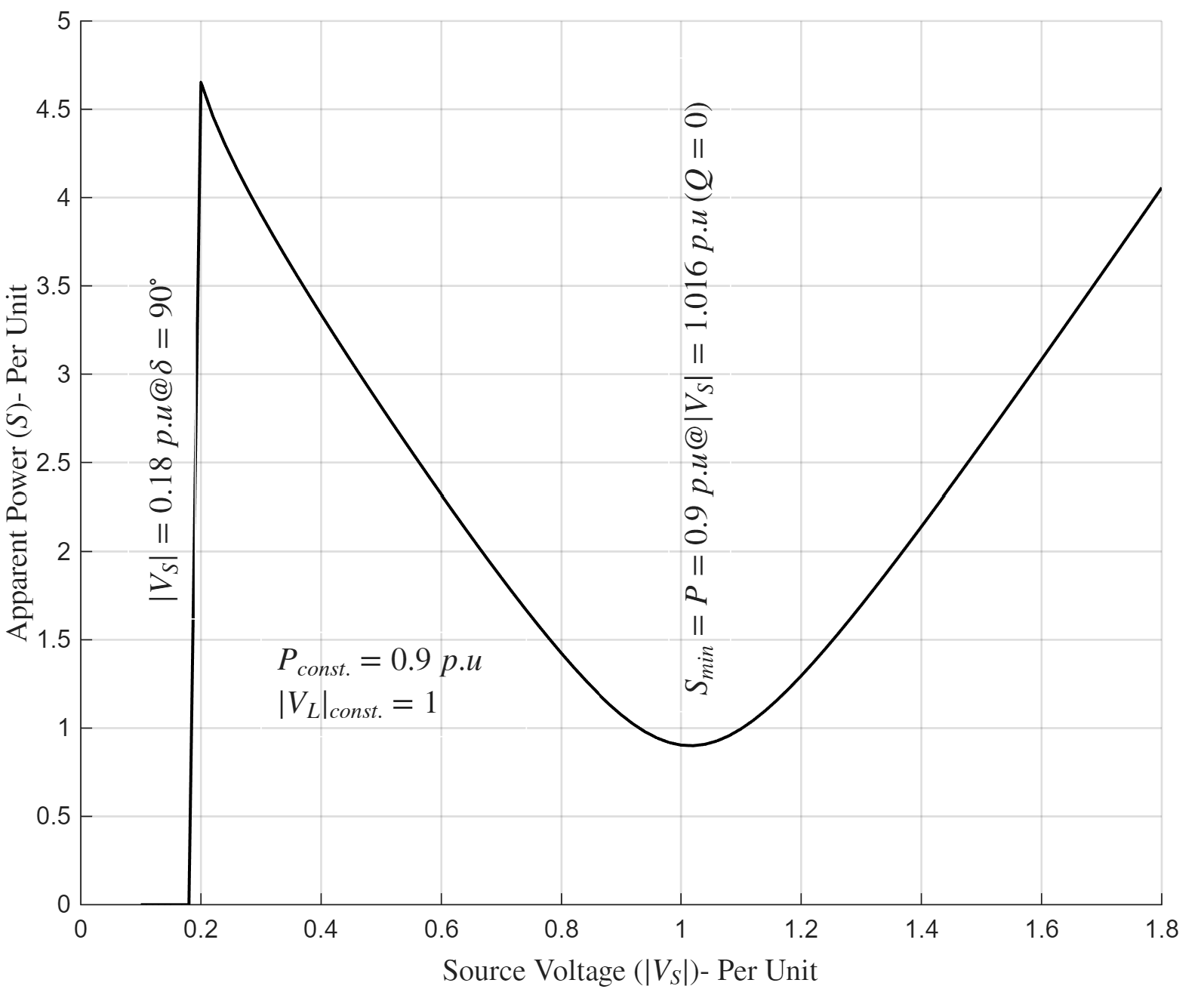}}
\caption{Apparent Power Versus $|V_{S}|$ at Constant $P$ and $V_{L}.$}
\label{fig:vcurve}
\end{figure}

To better understand the limits by which reactive power may be effective to compensate for a reduced $V_{S}$ we can use equations (\ref{eqP}) and (\ref{eqQ}) to plot the total apparent power~($S$) at the load for constant $V_{L}$ and $P$ when $|V_{S}|$, and consequently $Q$, vary. The apparent power can be obtained from $S^2=P^2+Q^2$, which provides a measure of how much total $VA$, and hence how much current at constant $V_{L}$, is required to meet the constant $P$ at various levels of $|V_{S}|$. Figure~\ref{fig:vcurve} shows a plot of total $S$ at the load for $0.1\le|V_{S}|\le1.8$ per unit for the same set up of Figure~\ref{fig:sld}. We saw that for $|V_{S}|<0.18$ the load voltage cannot be maintained due to theoretical limit at $\delta=90^\circ$, hence the graph cuts off at that voltage. The minimum $S$ occurs at $|V_{S}|\approx1.01~p.u$, where $Q=0$, which can also be seen in Figure~\ref{fig:vcurve}. For any smaller or larger $|V_{S}|$ a greater total $S$, and hence a larger current, is required.

In practice, thermal and electrical ratings of the substation equipment as well as the protection settings limit the amount by which current can increase to compensate for the reduced $|V_{S}|$. Therefore, there are both conceptual and practical limitations to reactive power's ability to help with load voltage ride-through. From a conceptual perspective, there is a limit to the amount of voltage dip that may be compensated by reactive power that occurs at the intersection of the load line and the horizontal axis, where $\delta=90^\circ$. From a practical standpoint, even when a voltage dip can be theoretically compensated by $Q$ it is likely that the substation equipment ratings, grid's capacity constraints, and system protection requirements limit the use of $Q$ for voltage ride-through. Practical considerations are often more limiting than the theoretical ones, as equipment ratings and protection requirements often do not allow much deviation. This leads us to an approach in which both $P$ and $Q$ may be used to address the problem of voltage ride-through for loads.

\section{Load Voltage Ride-Through: A Dual \emph{PQ} Approach}
With the limitation in using reactive power to address large load voltage dips the role of active power, $P$, in helping with voltage ride-through is investigated in this section. Circle diagrams can again help us for a deeper insight. Considering substation's equipment rating and protection requirements we may come up with a maximum current that can be safely, but temporarily, allowed at the load. This current would be the result of multiple practical constraints in the substation and  power distribution infrastructure serving the load. In this study we are keeping the load voltage at its constant rated value, hence the maximum current. $I_{max}$, may also be represented by a maximum apparent power, $S_{max}= |V_{L}||I_{max}|$. Since $S_{max}^2=P^2+Q^2$, the maximum apparent power constitutes a circle in the $PQ$ plane that defines the area of the maximum allowed total $S$ at the load.

The circle representing $S_{max}$ is shown in Figure~\ref{fig:pcircle} with its center at the origin and a radius of $S_{max}$. The rated apparent power of the substation is $1~p.u$. Let's assume that the practical limit for the temporary $S_{max}$ is $30\%$ higher than the rated $S$, i.e., $S_{max}=1.3~p.u$. Since the same power is supplied to the load $S_{max}$ is basically a measure of the maximum reactive power that may be added at the load. The source of this reactive power is not the load itself; it is very often provided by other auxiliary sources of $Q$. The area inside the $S$ circle is all the $P$ and $Q$ values that is allowed from the grid; any combination of $P$ and $Q$ resulting in an apparent power outside this circle must be avoided.

Although for a constant load power $P_{load}=0.9~p.u$, a constant $|V_{L}|=1$ may theoretically be maintained up to a reduced $|V_{S}|=0.18~p.u$, the practical limits imposed by the $S_{max}$ circle allows a minimum $|V_{S}|$ of about $0.83~p.u$ only. This can be seen at the intersection of the power line of $P=0.9~p.u$ and the circle of $S_{max}$. For any $|V_{S}|$ below this value maintaining a unity $V_{L}$ will require an apparent power the exceeds $S_{max}$.

These results and observations can also be obtained directly from the equations that are the basis of the plots in Figure~\ref{fig:pcircle}. This includes equation (\ref{eq:eqcirc}) and also the fact that $S_{max}^2=P^2+Q^2$ defines the values that are allowed for $P$ and $Q$. With the known $S_{max}$ and load power, $P$, we can obtain $Q$ for the maximum $S$, and then using equation (\ref{eq:eqcirc}) the associated $|V_{S}|$ is calculated. This is the minimum $|V_{S}|$ for which the load voltage can still be maintained at its rated value without exceeding $S_{max}$. For our system of $S_{max}=1.3~p.u$, $P=0.9~p.u$, $|V_{L}|=1$, and $X=0.2~p.u$, reactive power will be $Q=-0.94~p.u$, corresponding to $|V_{S}|=0.83~p.u$.

\subsection{\emph {Active Power: A New Dimension in Voltage Ride-Through}}\label{activepower}
An aspect of using reactive power in maintaining a constant voltage at the load is that once we are at a non-unity power factor the total current (or $S$) at the load is always higher than that of a unity power factor. In other words, maintaining the load voltage with the help of $Q$ during a dip in $|V_{S}|$ imposes an additional burden on the grid. Higher $Q$ demands higher total current in the form of apparent power, $S$, even though the active power is still the same, hence the practical limit in using $Q$ to maintain the load voltage.

In cases where the intersection of the line of the load power and a certain power circle is outside of the $S$ circle the active power of the load may also need to be partially supplied by other non-grid resources, such that the total $S$ from the grid remains below $S_{max}$, if at all possible, as we are going to see. The magnitude of the non-grid active power depends on the extend of the dip in $|V_{S}|$ as well as the power demand of the load.

For instance, in Figure~\ref{fig:pcircle} we observe that for $|V_{S}|=0.8$, the intersection with the line of $P=0.9$ is outside of the $S$ circle. For the $P$ line to be inside the $S$ circle it will need to be moved to the left. This means that the power supplied by the grid will be reduced, and that the difference between the load demand of $P_{load}=0.9$ and the new vertical $P$ line must be supplied by a non-grid resource. For the case shown in Figure~\ref{fig:pcircle}, if the new vertical line is at $P=0.5$, which is inside the $S$ circle for $|V_{S}|=0.8$, the power difference of $\Delta P=0.4~p.u$ will need to be supplied by other resources, such as a Battery Storage Energy System~(BESS). Note that sources of pure reactive power, such as STATCOMs or synchronous condensers, cannot meet the combined $P$ and $Q$ demand.

However, at higher dips in $|V_{S}|$ eventually we encounter situations where the power circle associated with a reduced $|V_{S}|$ is entirely outside of the $S$ circle at all times, i.e., regardless of how much active power is supplied by a non-grid resource the intersection of a vertical $P$ line and that power circle may not be located inside the circle of $S_{max}$, as we can see for power circles associated with $|V_{S}|=0.7$ or smaller. The reason is that $|V_{S}|$ has dipped so much that just maintaining $V_{L}$ at unity will require levels of reactive power that alone exceeds the permitted $S_{max}$.

The minimum voltage, $|V_{S}|_{min}$, beyond which the power circle lies completely outside the $S$ circle can be calculated from equation (\ref{eq:eqcirc}). This voltage is associated with the situation where the entire apparent power from the grid is in the form of reactive power to maintain the load voltage, i.e., when $Q=-S_{max}$ while $P=0$ for $|V_{L}|=1$. For our system with $X=0.2~p.u$ and $S_{max}=1.3~p.u$ the calculation provides $|V_{S}|_{min}=0.74~p.u$. This is the intersection of the lower side of the $S$ circle and the $Q$ axis where $P=0$. In other words, if $|V_{S}|$ drops to lower than $0.74~p.u$ the load voltage can no longer be maintained at the unity level without exceeding the maximum permitted $S_{max}$. 

In general, we can find the minimum source voltage, $|V_{S}|_{min}$, for which the load voltage may be maintained at unity, using equation (\ref{eq:eqcirc}). Considering that $Q=-S_{max}$, $P=0$, and $V_{L}=1$, the general expression for $|V_{S}|_{min}$ is obtained as:
\begin{equation}
|V_{S}|_{min}=1-XS_{max}\hspace{1mm}p.u \label{eq:vmin}
\end{equation}
Note that $|V_{S}|_{min}$ is a function of system's $X$ and $S_{max}$ only, independent from $P$, as expected. Both $X$ and $S_{max}$ are characteristics of the substation and they can vary case by case. We can also observe that a higher $S_{max}$ results in a smaller $|V_{S}|_{min}$, i.e., it will be possible to maintain a unity load voltage with a smaller $|V_{S}|$. This of course will require higher reactive power, hence higher current, at the load. With $X=0.2~p.u$ and $S_{max}=1.3~p.u$ we get the same $|V_{S}|=0.74~p.u$ from equation (\ref{eq:vmin}).

On the other hand, the intersection of the load line, $P=P_{load}$, and the $S$ circle is the minimum $|V_{S}|$ that load voltage can still be maintained at unity without a need for active power compensation by non-grid resources. For all the source voltages that are between $|V_{S}|_{m}$ and $|V_{S}|_{min}$ the load voltage can be maintained at unity during a dip in $|V_{S}|$ only if active power is also supplied to the load by non-grid resources. $|V_{S}|_{m}$ can be calculated by knowing $P_{load}$, $S_{max}$, and system's $X$ while $|V_{L}|=1~p.u$ . Obtaining $Q$~(negative value) from $S_{max}^2=P_{load}^2+Q^2$ and then using equation (\ref{eq:eqcirc}) we can calculate $|V_{S}|_{m}$ as:
\begin{equation}
|V_{S}|_{m}^{2}=1+X^{2}S_{max}^2-2X\sqrt{S_{max}^2-P_{load}^2}\hspace{2mm}p.u \label{eq:VM}
\end{equation}
Note that $|V_{s}|_{m}$, in addition to $X$ and $S_{max}$, also depends on $P_{load}$, unlike $|V_{S}|_{min}$ which is independent from load. With $P_{load}=0.9~p.u$ the calculated $|V_{S}|_{min}$ is about $0.83~p.u$, i.e., as long as the source voltage is above $0.83$, and for $P_{load}=0.9~p.u$, load voltage can still be maintained by just reactive power compensation with no need for non-grid active power.

When the dipped grid voltage is between $|V_{S}|_{m}$ and $|V_{S}|_{min}$, while the vertical $P$ line is outside of the $S$ circle, we can calculate the power from the non-grid resource, $P_{vrt}$, that is needed for voltage ride-through. From the circle diagrams in Figure~\ref{fig:pcircle}, the new operating region associated with a power circle will be between two points: the intersection of the power circle with the $S$ circle on the one side, and the intersection of the same power circle with the $Q$ axis. This part of the power circle is located inside the $S$ circle. The reactive power at the intersection of the power circle with the $S$ circle can be obtained from equation (\ref{eq:eqcirc}) and the equation of the $S$ circle. This calculation provides the reactive power at this intersection, $Q_{ints.}$, as:
\begin{equation}
Q_{ints.}=\frac{1}{2X}\left(|V_{S}|^2-|V_{L}|^2-\frac{X^{2}S_{max}^2}{|V_{L}|^2}\right)\hspace{2mm}p.u \label{eq:qint}
\end{equation}
The reactive power at the intersection of the power circle with the $Q$ axis, which is the minimum reactive power~(absolute value) on the power circle, can be calculated form equation (\ref{eq:eqcirc}) by having $P=0$. This leads to the reactive power at this intersection, $Q_{min}$, as:
\begin{equation}
Q_{min}=\frac{|V_{L}|}{X}\left(|V_{S}|-|V_{L}|\right)\hspace{2mm}p.u \label{eq:qmin}
\end{equation}
Both $Q_{ints.}$ and $Q_{min}$ are negative values for leading currents. We will pick a value for the reactive power at the load that is between the values of these two reactive powers, i.e., $|Q_{min}|<|Q|<|Q_{ints.}|$. Considering the limited resources, this may be optimized by selecting a smaller absolute value for $Q$. Then the new grid power, $P_{grid}$, is calculated from equation (\ref{eq:eqcirc}) from the $Q$ that we have picked. This completes the calculation of the new active and reactive power from the grid, associated with a dipped grid voltage where its power circle is located inside the $S$ circle. The active power from the non-grid resource is then simply the difference between the new grid power and the original load power, i.e., $P_{vrt}=\left(P_{load}-P_{grid}\right)$.

\subsection{\emph {An Implementation Procedure for the Dual PQ Approach}}
Based on the relative magnitude of the source voltage, $|V_{S}|$, three scenarios may be considered for implementation of the dual $PQ$ method for load voltage ride-though. There are two thresholds for $|V_{S}|$ that determine the type of response to a voltage dip during a grid disturbance: the voltage $|V_{S}|_{m}$ that provides the margin where reactive power alone can help the load ride through the disturbance, and the minimum voltage, $|V_{S}|_{min}$, beyond which load voltage may not be held at its unity value with any amount of reactive or active power compensation. For the case where $|V_{S}|_{min}<|V_{S}|<|V_{S}|_{m}$ supply of both $P$ and $Q$ is needed to maintain the load voltage at unity. Values of $|V_{S}|_{min}$ and $|V_{S}|_{m}$ are calculated from equations (\ref{eq:vmin}) and \ref{eq:VM}, respectively.

\begin{itemize}
\item \emph{Case 1- $|V_{S}|\ge|V_{S}|_{m}$:} With $|V_{L}|=1$, $P=P_{load}$, dipped $|V_{S}|$ and system's $X$, equation (\ref{eq:eqcirc}) can be used to calculate the amount of the total reactive power, $Q$~(negative), that is needed to maintain the constant load voltage. The non-grid resource will then respond by providing $Q$ at the load, considering a unity load voltage.

For our example system with $|V_{S}|_{m}=0.83$ the total reactive power at the load is calculated as $Q=-0.79$ when $|V_{S}|$ drops to $0.86~p.u$. Of course, when the grid goes back to its normal conditions the additional reactive power towards $Q$ will no longer be needed from the non-grid resource.

To put this into perspective, note to the relatively large amount of reactive power, $79\%$ of the base apparent power of the large load, that is needed to maintain a unity load voltage for a $14\%$ dip in the grid's voltage. This example provides a measure of the scale of the resources required for voltage ride-through.

\item \emph{Case 2- $|V_{S}|_{min}<|V_{S}|<|V_{S}|_{m}$}: This is when both $P$ and $Q$ are needed from the non-grid resource to maintain the load voltage. The calculations for active and reactive power required from the non-grid resource, $P_{vrt}$ and $Q$, were described in section \ref{activepower}.

For our example system, when $|V_{S}|$ drops to $0.76~p.u$, calculations provide $Q_{ints.}=-1.225~p.u$, and $Q_{min}=-1.2~p.u$. The new reactive power at the load is chosen to be $Q=-1.21~p.u$, from which the new power from the grid is calculated as $P_{grid}=0.275~p.u$. The difference in power, $\Delta P=P_{load}-P_{grid}$ is what needs to be provided by the non-grid resource, i.e., $P_{vrt}=0.625~p.u$. Both $Q$ and $P_{vrt}$ need to be provided by the non-grid resource. With this large voltage dip, a significant reactive power is required to maintain the voltage which limits the power from the grid, hence a large active power of $0.625~p.u$ is required from the non-grid resource.

From the active and reactive power the total apparent power to be provided by the non-grid resource is $S=1.36~p.u$. This is almost the same as $S_{max}$ and shows that large non-grid resources will be needed to meet the voltage ride-through requirements at significant voltage dips.

\item \emph{Case 3- $|V_{S}|\le|V_{S}|_{min}$:} This is a case when the amount of reactive power alone required to maintain the load voltage exceeds $S_{max}$. Therefore, the load may need to disconnect from the grid as the load voltage cannot be maintained. The next available options depend on the severity of the voltage dip. One option is to verify if it is possible to maintain the load voltage at a lower, but still acceptable, level, e.g., at $0.9~p.u$ instead of unity. This can quickly turn into a situation where the entire load power needs to be provided by the non-grid resource while it also needs to provide some reactive power to maintain load voltage at an acceptable level. While this may technically be achievable the practical and cost consequences may not be justified.

The other option, perhaps more practical, is for the large load to disconnect from the grid during a severe disturbance and voltage dip while the emergency resources continue providing power to the load. When the grid conditions are back to normal, i.e., when $|V_{S}|$ is at an acceptable level, and to help with maintaining stability of the grid, either the large load, or an alternative load such as a charging BESS, may be quickly connected back to the grid, within a time specified by the grid stability requirements. This prevents the grid from experiencing sudden loss of large amounts of load, which can then trigger grid stability problems.
\end{itemize}
It was illustrated that enabling a broader voltage ride-through capability for large loads requires non-grid resources with the ability to supply both active and reactive power. Response time of these resources should be fast enough so that loads are not transferred to emergency resources or otherwise disconnected. In AI data centers these emergency resources typically consist of UPS systems that transfer the load to emergency within programmable time and voltage settings~\cite{chen}. UPS systems have physical characteristics and limitations that allow them to operate only within a narrow window of grid voltage before they switch to their emergency resources, e.g., within $\pm15\%$ of the rated input voltage. For voltages below the lower limit, either the load must be proportionally reduced to prevent switching to emergency mode, which is often not practical, or UPS will need to switch to emergency mode to be able to supply the full load, resulting the load being removed from the grid. Therefore, the voltage ride-through response should be designed to prevent the large load from being disconnected from the grid, if and when possible, depending on the extend of the grid disturbance and voltage dip. While UPS voltage and time settings for switching to emergency mode may be adjustable, under severe voltage dips protection and current limitation concerns can take priority and switch the UPS to emergency mode, overriding other settings.

Examples of non-grid resources that could provide the required response time as well as flexibility to supply variable active and reactive power are BESS and e-STATCOM, the latter typically with supercapacitors as the source of energy. One distinction is that the energy density of supercapacitors is much lower than BESS, and this can create challenges for e-STATCOM to meet energy needs during voltage ride-through. Synchronous condensers are traditionally used to provide $Q$ for voltage compensation within small voltage drops during normal conditions of the grid, while they are not designed to provide active power. Besides, synchronous condensers are sensitive to large transients on the grid and may not be able to provide their full capacity during transients. Eventually, the specific large load characteristics and grid requirements determine the type of technology that can be used for voltage ride-through purposes.

Finally, there are levels of voltage dip in the power grid that make it impossible or impractical to maintain an acceptable on-grid load voltage, most often due to current rating limitations in the power infrastructure of large loads. This is a situation where disconnecting the load from the grid may become inevitable, i.e., switching the IT loads to UPS units. Hence, helping with the grid stability may need the attention to be shifted to how quickly a large load can return on the grid when the grid is back to normal conditions. The sooner a large load returns on the grid the lower the risk to grid stability. This may be achieved by quick return of the large load itself on the grid, or alternatively, by emulating the load by other means, such as a charging BESS. \cite{cui} provides an analytical framework for safe reconnection of large loads to the grid.

Each of the above scenarios has its complexity. In the former, the large load itself needs to return on the grid within a short time post-transient, a period that can be specified by stability studies, generally within a few short seconds. One challenge is that UPS units are traditionally designed to wait for double-digit seconds to ensure the health of the grid before switching back to the normal grid source, which is likely beyond the stability requirements of the grid. This may be addressed by design changes to shorten the UPS wait time. Additionally, only IT loads of data centers are protected by UPS. Non-IT loads, particularly the large cooling systems, may have other equipment, such as Variable Frequency Drives (VFD), that are also sensitive to voltage transients, disconnecting from the source upon voltage dips. Re-starting VFDs can also require a wait time that may be longer than grid stability needs, which demands another level of re-engineering to ensure quick re-starting time. On the other hand, the latter scenario of emulating the large load with a charging BESS may have the advantage of bypassing the aforementioned UPS and VFD problems with quick return to normal source, but the required size and complexity of the scheme can pose considerable planning, cost, and technical challenges.

\section{Conclusion}
AI data centers are large loads with significant size and complexity; they are hundreds of megawatts or even over one gigawatt in size. At this scale, they can pose new challenges to the power grid, including a sudden disconnection from the grid due to insufficient voltage ride-through capability that can create stability risks.

There are significant research gaps in analytical understanding that is required to address the problem of voltage ride-through in large loads. While there have been efforts in the industry to simulate specific project-based scenarios to study the impacts and risks associated with insufficient voltage ride-through capabilities in AI data centers, the studies often lack mathematical analysis and theoretical structure to generate a true and new knowledge that is needed to tackle this problem. This has led to some level of uncertainty not only in a comprehensive technical understanding of the problem but also in evaluating effective solutions as the conventional reactive power compensation techniques may not be adequate. The theoretical investigation presented in this paper provides a technical framework to better understand the problem of voltage ride-through in large loads, how to define the effective solutions, and what theoretical and practical limitations should be expected from these solutions. This can ultimately inform industry standards and regulations, as well as utility grid planning to address potential reliability and stability risks associated with large loads.

While reactive power compensation has been historically used to regulate voltage during normal operation conditions it was shown that the problem of voltage ride-through in large loads is of a different nature, where the goal of the reactive power compensation is not the conventional power factor correction, but is to use the dynamics of the interrelation between $V$, $Q$, and $P$ to maintain the load voltage at an acceptable level during transient voltage dip events, such as faults, on the power grid. This provides a novel conceptual approach to reactive power where it is not viewed just as a tool for the narrow purpose of small-scale voltage regulation but in its full correlation with $P$ and $V$ where both $Q$ and $P$ may be adjusted over a wide range, with the help of non-grid resources, to maintain an acceptable load voltage during \emph{transient} events. It was illustrated that while it is theoretically feasible to use reactive power to maintain the load voltage even at significant voltage dips of over $50\%$ the practical ratings and limitations of the power infrastructure of large loads often allow only a limited use of reactive power, before the maximum current ratings are reached.

It was also shown that using active power ($P$), supplied by a non-grid resource, can extend the degree by which load voltage may be maintained during grid disturbances. This new concept provides an approach where both $P$ and $Q$ are used to help address the problem of voltage ride-through in large loads. This will require non-grid resources that are capable of providing both $P$ and $Q$, and fast enough in response to grid's transient events, typically lasting only for several power system cycles.

The analysis and results in Section III furnish the basis to specify the active and reactive power capabilities, i.e., sizing, of the resources needed to enable a large load to ride through voltage disturbances, within defined limits, based on the equivalent reactance of the load's substation, $X$, and the maximum short term MVA rating of the substation, $S_{max}$. Ultimately, when there is a significant voltage dip in the power grid it may not be practical, or even technically possible, to maintain an acceptable load voltage. To protect the IT loads and prevent disruption in their operation the UPS units will transfer the load to their emergency source, which means removing the IT load from the grid. This scenario will require very fast return of UPS units to the grid after the grid is back to normal conditions, which pose technical difficulties with the existing UPS and VFD designs with relatively high wait times before reconnecting to normal grid source.


\end{document}